\documentstyle[12pt]{article}
\huge
 
\def\setzero{\setcounter{equation}{0}}

\newcounter{eqalph}

\begin{document}

\baselineskip 18pt
\newcommand{\ii}{\mbox{i}}
\def \sech{{\rm sech}}
\def \tanh{{\rm tanh}}
\def \cn{{\rm cn}}
\def \sn{{\rm sn}}
\def\bm#1{\mbox{\boldmath $#1$}}
\newfont{\bg}{cmr10 scaled\magstep4}
\newcommand{\bzr}{\smash{\hbox{\bg 0}}}
\newcommand{\bzl}{%
   \smash{\lower1.7ex\hbox{\bg 0}}}
\title{The  Davey-Stewartson  system 
\\ 
and
\\ 
the B\"{a}cklund Transformations}
\date{\today}

\vspace{0.5cm}
\author{  
\vspace{0.5cm}
Masato {\sc Hisakado}
\\
{\small\it Department of Pure and Applied Sciences,}\\
{\small\it University of Tokyo,}\\
{\small\it 3-8-1 Komaba, Meguro-ku, Tokyo, 153, Japan}
\\
}
\maketitle

\vspace{20 mm}

Abstract

We  consider the (coupled) Davey-Stewartson (DS)  system and 
its B\"{a}cklund transformations (BT).
Relations among   the DS   system, the double 
Kadomtsev-Petviashvili  (KP) system  and 
 the Ablowitz-Ladik hierarchy (ALH)  are  established.
The   DS  system  and the double KP system
are equivalent.
The ALH is the BT of  the DS  system 
in  a certain reduction. 
 {From} the BT of coupled  DS system we  can obtain   new coupled 
derivative nonlinear  Schr\"{o}dinger equations.


\vfill
\par\noindent
{\bf Key Words :}
Davey-Stewartson system, KP system, Ablowitz-Ladik hierarchy,
coupled DNLS equations

\newpage

\section{Introduction}

The Davey-Stewartson  (DS) equation is one of a few 
integrable equations  in multi-dimensions
which have significances, and  has received 
considerable attention during the last decade.\cite{ZM}
The DS equation has a new type of solution 
which is called dromion.\cite{DS1}, \cite{DS2}
The dromion  has the remarkable property that 
it decays exponentially  in two spatial dimensions.
It is well known that the DS equation is the simplest 
nontrivial equation in the two-component 
Kadomtsev-Petviashvili (KP)  equation.\cite{Bo}

Recently a new infinite set of commuting 
``ghost''  symmetries was proposed 
for the KP-type integrable hierarchy by Aratyn, Nissimov 
and Paceva.\cite{Ar}
These symmetries  allow for a Lax representation 
in which the hierarchy is  realized as standard isospectral flows.
This gives rise to a new double KP hierarchy 
embedding ghost and original KP type Lax hierarchies 
connected to each other by a duality mapping.


 The 
universality of the Ablowitz-Ladik 
hierarchy (ALH)
was pointed out by V.E.Vekslerchik.\cite{Ve}
The ALH ``contains'' the 2D Toda lattice,
DS equation and KP equation.\cite{Ve1},\cite{Ve2}(See \cite{MH1}.)
In \cite{Ve1} the   
several solutions 
were obtained using the fact that  
the solutions of the ALH satisfy  
DS equation.


In one component case 
there are two derivative nonlinear Schr\"{o}dinger (DNLS) equations.
One is Kaup-Newell (KN) equation   and 
the other is  Chen-Lee-Liu (CLL) equation.\cite{DNLS1},\cite{DNLS2}
It is well known that 
these two equations are gauge equivalent.
In the previous paper 
we discuss the coupled DNLS equations
which are the coupled version of KN equation.\cite{h2}

In this paper 
we  consider the (coupled) DS system and 
its B\"{a}cklund transformations (BT).
Relations   among  the DS   system, the double KP system  and 
 the ALH are  established.
The  DS  system  and the double KP system
are equivalent. 
The ALH is the BT of  the DS  system 
in the case $t_{k}=\bar{t}_{k}^{*}$ and  $u_{k}=v_{k}^{*}$ 
where the asterisk  is   conjugation.
 {From} the BT of the coupled DS system we  can obtain   new coupled 
DNLS equations which are  the coupled type of CLL equation. 
This suggests that
in the multi-component case 
there are two types
of coupled DNLS equations 
which are not gauge equivalent.

The paper is organized as follows.
In  section 2 
we obtain  the Lax pair of the coupled DS equations.
This  is a  new integrable  system.
In  section 3 
we construct the BT of the coupled DS equations.
In  section 4 
we establish a  
relation between
the BT of the coupled DS equation  and the ALH.
In  section 5 
 we consider a  relation between 
the double KP system and the DS system.
In  section 6
we  get  new coupled DNLS equations
using  the BT which is obtained in the section 2.
The last section is devoted to the 
concluding remarks.

\setzero

\section{Coupled Davey-Stewartson system}
 
We consider the Lax representation of spatially two-dimensional 
 systems.We denote the time by $T$ and the space coordinates
 by $X$ and $Y$,
As the Lax representation let 
\begin{equation}
[L, \frac{\partial }{\partial T}-A]=0,
\label{lax}
\end{equation}
where
\begin{eqnarray}
L=
\alpha\frac{\partial }{\partial Y}+
\left(
\begin{array}{ccc}
l+1& 0&0 \\
0&l
&0\\
0& 0& l
\end{array}
\right)
\frac{\partial }{\partial X}
+
\left(
\begin{array}{ccc}
0& \pm v^{(1)}&\pm v^{(2)} \\
 u^{(1)}&0
&0\\
 u^{(2)}& 0& 0
\end{array}
\right),
\label{L}
\end{eqnarray}
and 
\begin{eqnarray}
A=
\left(
\begin{array}{ccc}
a+1& 0&0 \\
0&a
&0\\
0& 0& a
\end{array}
\right)
\frac{\partial^{2} }{\partial X^{2}}
+
\left(
\begin{array}{ccc}
0& \pm v^{(1)}&\pm v^{(2)} \\
u^{(1)}&0
&0\\
u^{(2)}& 0& 0
\end{array}
\right)
\frac{\partial }{\partial X}
+
\left(
\begin{array}{ccc}
r_{11}& \tilde{\varphi}_{1}&\tilde{\varphi}_{2} \\
\varphi_{1}&r_{22}
&r_{32}\\
\varphi_{2}& r_{23}& r_{33}
\end{array}
\right).
\label{A}
\end{eqnarray}
Here 
\begin{eqnarray}
\varphi_{i}&=&-\alpha\frac{\partial u^{(i)}}{\partial Y}+
(2a-l+1)\frac{\partial u^{(i)}}{\partial X},
\nonumber \\
\tilde{\varphi}_{i}&=&
\pm
[\alpha\frac{\partial v^{(i)}}{\partial Y}
+(l-2a)\frac{\partial v^{(i)}}{\partial X}],
\;\;\;i=1,2.
\end{eqnarray}
To simplify the expressions, we introduce
\begin{eqnarray}
\tilde{D}_{1}&=&\alpha\frac{\partial}{\partial Y}+l\frac{\partial}{\partial X},
\;\;\;
\tilde{D}_{2}=\alpha\frac{\partial}{\partial Y}+(l+1)
\frac{\partial}{\partial X},
\nonumber \\
\tilde{D}_{3}&=&\alpha\frac{\partial}{\partial Y}+
(l-2a)\frac{\partial}{\partial X},
\;\;\;
\tilde{D}_{4}=\alpha\frac{\partial}{\partial Y}+
(l-2a-1)\frac{\partial}{\partial X},
\nonumber \\
2D_{1}&=&\tilde{D}_{1}\tilde{D}_{4}+\tilde{D}_{2}\tilde{D}_{3},\;\;\;
D_{2}=\tilde{D}_{1}\tilde{D}_{2},\;\;\;
D_{3}=\tilde{D}_{1}\tilde{D}_{4}.
\label{2.5}
\end{eqnarray}
And the conditions (\ref{lax}) with  (\ref{L})-(\ref{2.5}) are 
reduced to a  system of equations
\begin{eqnarray}
& &
\frac{\partial u^{(1)}}{\partial T}
=D_{1}u^{(1)}+p^{(1)}u^{(2)}-r_{23}u^{(2)},\;\;\;
\nonumber \\
& &
\frac{\partial u^{(2)}}{\partial T}
=D_{1}\phi_{1}+p^{(2)}u^{(2)}-r_{32}u^{(1)},
\label{DS1}
\end{eqnarray} 
where $p^{(1)}=r_{11}-r_{22}$,  $p^{(2)}=r_{11}-r_{33}$ 
and 
\begin{eqnarray}
D_{2}p^{(1)}&=&\mp (2D_{1}u^{(1)}v^{(1)}+D_{3}u^{(1)}v^{(2)}),
\nonumber \\
D_{2}p^{(2)}&=&\mp (2D_{1}u^{(1)}v^{(2)}+D_{3}u^{(1)}v^{(1)}),
\nonumber \\
\tilde{D}_{1}r_{32}&=&\pm\tilde{D}_{3}u^{(2)}v^{(1)},\;\;\;
\tilde{D}_{1}r_{23}=\pm\tilde{D}_{3}u^{(1)}v^{(2)}.
\label{DS2}
\end{eqnarray}

The particular  cases come from 
 the degenerations  of the quadratic form of 
$D_{1}$. Let $a=0$ then the system reduces to 
\begin{eqnarray}
& &
\frac{\partial  }{\partial t}u^{(1)}
=
\frac{\partial^{2}}{\partial x^{2}}u^{(1)}
+
p^{(1)}\phi_{1}-r_{23}u^{(2)},
\;\;\;\;
\frac{\partial  }{\partial t}u^{(2)}
=
\frac{\partial^{2}}{\partial x^{2}}u^{(2)}
+
p^{(2)}u^{(2)}-r_{32}u^{(1)},
\nonumber \\
& &
\frac{\partial }{\partial y}(p^{(1)}\pm u^{(2)}v^{(2)})=\mp 2
\frac{\partial }{\partial x}(u^{(1)}v^{(1)}),
\;\;\;\;
\frac{\partial }{\partial y}(p^{(2)}\pm  u^{(1)}v^{(1)})=\mp 2
\frac{\partial }{\partial x}(u^{(2)}v^{(2)}),
\nonumber \\
& &
\frac{\partial }{\partial y}r_{32}=\pm \frac{\partial }{\partial y}
(u^{(2)}v^{(1)}),\;\;\;\;
\frac{\partial }{\partial y}r_{23}=\pm \frac{\partial }{\partial y}
(u^{(1)}v^{(2)}),
\label{ds1}
\end{eqnarray}
where
\begin{equation}
\frac{\partial }{\partial y}=\alpha\frac{\partial }{\partial Y}
+(l+1)\frac{\partial }{\partial X},\;\;\;
\frac{\partial }{\partial x}=\alpha\frac{\partial }{\partial Y}
+l\frac{\partial }{\partial X},
\end{equation}
and we rewrite $T=t$. 

Let $a=-1$ then the system reduces to
\begin{eqnarray}
& &
\frac{\partial  }{\partial \tau}u^{(1)}
=
\frac{\partial^{2}}{\partial y^2}u^{(1)}
+
q^{(1)}u^{(1)}-\tilde{r}_{23}u^{(2)},
\;\;\;\;
\frac{\partial  }{\partial \tau}u^{(2)}
=
\frac{\partial^{2}}{\partial y}u^{(2)}
+
q^{(2)}u^{(2)}-\tilde{r}_{32}u^{(1)},
\nonumber \\
& &
\frac{\partial }{\partial x}(q^{(1)}\pm u^{(2)}v^{(2)})=\mp 2
\frac{\partial }{\partial y}(u^{(1)}v^{(1)}),
\;\;\;\;
\frac{\partial }{\partial x}(q^{(2)}\pm u^{(1)}v^{(1)})=\mp 2
\frac{\partial }{\partial y}(u^{(2)}v^{(2)}),
\nonumber \\
& &
\frac{\partial }{\partial y}\tilde{r}_{32}=\pm (2\frac{\partial }{\partial x}
-\frac{\partial }{\partial y})
(u^{(2)}v^{(1)}),\;\;\;
\frac{\partial }{\partial y}\tilde{r}_{23}=\pm (2\frac{\partial }{\partial x}
-\frac{\partial }{\partial y})
(u^{(1)}v^{(2)}),
\label{ds2}
\end{eqnarray}
where $\tau=T$, $q_{1}=p_{1}$ and $q_{1}=p_{1}$  .

In the one component case ($u^{(2)}=v^{(2)}=0$)
 (\ref{ds1}) and (\ref{ds2}) become
\begin{equation}
\frac{\partial }{\partial t}u
=\frac{\partial^{2}}{\partial x^{2}}u+2pu,
\;\;\;
\frac{\partial}{\partial y}p=\mp\frac{\partial}{\partial x}(uv),
\label{dd1}
\end{equation}
and 
\begin{equation}
\frac{\partial }{\partial \tau}u
=\frac{\partial^{2}}{\partial y^{2}}u+2qu,
\;\;\;
\frac{\partial}{\partial x}q=\mp\frac{\partial}{\partial y}(uv),
\label{dd2}
\end{equation}
where $2p=p^{(1)}$ and $2q=q^{(1)}$.
(\ref{dd1}) and (\ref{dd2}) are  compatible and any linear combination 
of them is  integrable.
We call 
this system with higher flows the DS system.\cite{ZM}

In the (1+1) dimensional case ($ y =x$)
 (\ref{ds1}) and (\ref{ds2}) become 
the coupled nonlinear Sch\"{o}rdinger (NLS) equation.\cite{Ma}
\begin{equation}
\partial_{t}u^{(i)}=\partial_{x}^2 u^{(i)}
\mp 2(u^{(1)}v^{(1)}+u^{(2)}v^{(2)})u^{(i)},
\nonumber \\
-\partial_{t}v^{(i)}=\partial_{x}^2 v^{(i)}
\mp2(u^{(1)}v^{(1)}+u^{(2)}v^{(2)})v^{(i)},
\label{cnls}
\end{equation}
where $i=1,2$.

\section{B\"{a}cklund Transformations}

We shall consider a sequence $u^{(i)}_{n}$ and $v^{(i)}_{n}$
generated by the  auto-B\"{a}cklund transformation (BT).
In the operator form  BT
is defined by \cite{MY}
\begin{equation}
W_{n}L_{n}=L_{n+1}W_{n}.
\label{BT}
\end{equation}
An implicit the BT of (\ref{ds1}) and (\ref{ds2})
corresponds to 
\begin{eqnarray}
W_{n}=
\left(
\begin{array}{ccc}
1& -v^{(1)}_{n+1}&-v^{(2)}_{n+1} \\
u^{(1)}_{n}&\partial_{y}-\partial_{x}-w^{(11)}_{n}
&-w^{(21)}_{n}\\
u^{(2)}_{n}&-w_{n}^{(12)} & \partial_{y}-\partial_{x}-w^{(22)}_{n}
\end{array}
\right),
\label{W}
\end{eqnarray} 
Note that we only consider the bright case, that is,
the sign is ``$-$''  in (\ref{L}) and (\ref{A}).

{From}  (\ref{L}), (\ref{BT}) and (\ref{W})
we can obtain 
\begin{eqnarray}
-\partial_{x}u_{n}^{(1)}=u_{n+1}^{(1)}+
w_{n}^{(11)}u_{n}^{(1)}+w_{n}^{(21)}u_{n}^{(2)},
\nonumber \\
-\partial_{x}u_{n}^{(2)}=u_{n+1}^{(2)}+
w_{n}^{(22)}u_{n}^{(2)}+w_{n}^{(12)}u_{n}^{(1)},
\nonumber \\
\partial_{x}v_{n}^{(1)}=v_{n-1}^{(1)}+
w_{n-1}^{(11)}v_{n}^{(1)}+w_{n-1}^{(21)}v_{n}^{(2)},
\nonumber \\
\partial_{x}v_{n}^{(2)}=v_{n-1}^{(2)}+
w_{n-1}^{(22)}v_{n}^{(2)}+w_{n-1}^{(12)}v_{n}^{(1)},
\label{ce1}
\end{eqnarray}
and 

\begin{eqnarray}
\partial_{y}w_{n}^{(11)}=-u_{n+1}^{(1)}v_{n+1}^{(1)}
+u_{n}^{(1)}v_{n}^{(1)},
\nonumber \\
\partial_{y}w_{n}^{(22)}=-u_{n+1}^{(2)}v_{n+1}^{(2)}
+u_{n}^{(2)}v_{n}^{(2)},
\nonumber \\
\partial_{y}w_{n}^{(21)}=-u_{n+1}^{(2)}v_{n+1}^{(1)}
+u_{n}^{(2)}v_{n}^{(1)},
\nonumber \\
\partial_{y}w_{n}^{(11)}=-u_{n+1}^{(1)}v_{n+1}^{(2)}
+u_{n}^{(1)}v_{n}^{(2)},
\label{ce2}
\end{eqnarray}
Here we introduce  new nonlocal dynamical variables 
$w^{(ij)}_{n}$.

\setzero
\section{Ablowitz-Ladik Hierarchy }

We consider the one component case of (\ref{ce1}) and (\ref{ce2})
\begin{eqnarray}
-\partial_{x}u_{n}&=&u_{n+1}+u_{n}w_{n},
\nonumber \\
 \partial_{x}v_{n}&=&v_{n-1}+v_{n}w_{n-1},
\nonumber \\
 \partial_{y}w_{n}&=&\partial_{x}(u_{n}v_{n+1})=
u_{n}v_{n}-u_{n+1}v_{n+1},
\label{al2}
\end{eqnarray}

Here we set 
\begin{equation}
u_{n}=v_{n}^{*},\;\;\;x=y^{*}.
\label{cc}
\end{equation}
The asterisk  means  conjugation:
\begin{equation}
(f^{*})^{*}=f,
\end{equation}
where $f$ is an  arbitrary function.
{From} this constraint we can obtain the following equations

\begin{eqnarray}
\partial_{y}u_{n}&=&u_{n-1}+u_{n}w_{n-1}^{*},
\nonumber \\
- \partial_{y}v_{n}&=&v_{n+1}+v_{n}w_{n}^{*},
\nonumber \\
 \partial_{x}w_{n}^{*}&=&\partial_{y}(u_{n+1}v_{n})=
u_{n}v_{n}-u_{n+1}v_{n+1}.
\label{al1}
\end{eqnarray}
We can obtain also (\ref{al1}) by  an exchange of 
$u_{n}\leftrightarrow  v_{n}$ in (\ref{al2}).

{From} the last equations of (\ref{al2}) and (\ref{al1}), we have 
\begin{equation}
w_{n}=u_{n+1}v_{n}+C,\,\,\,
w_{n}^{*}=u_{n}v_{n+1}+C^{*},
\label{al3}
\end{equation}
where $C$ and $C^{*}$ are constants.
Hereafter for simplicity we set $C=C^{*}=0$.
(\ref{al2})-(\ref{al3}) are nothing but the Ablowitz-Ladik hierarchy (ALH).
Thus, the ALH  can be viewed as sequences of the 
BT  of the DS system.

In fact the second flow  of the ALH  are
\begin{equation}
\partial_{t}u_{n}=p_{n}p_{n+1}u_{n+2}
+p_{n}u_{n+1}v_{n-1}u_{n}
+p_{n}u_{n+1}^{2}v_{n},
\end{equation}
\begin{equation}
\partial_{t}v_{n}=-p_{n}p_{n-1}v_{n-2}
-p_{n}v_{n-1}^{2}u_{n}
-p_{n}v_{n-1}v_{n}u_{n+1},
\label{al4}
\end{equation}
where $p_{n}=1+u_{n}v_{n}$.

One can obtain straightforwardly \cite{Ve1}
\begin{equation}
\partial_{t}u_{n}=\partial_{x}^{2}u_{n}+2A_{n}u_{n},
\;\;\;A_{n}=p_{n}u_{n+1}v_{n-1}.
\label{dds1}
\end{equation}
On the other hand 
we can obtain
\begin{equation}
\partial_{y}A_{n}=p_{n}(-v_{n}u_{n+1}+v_{n-1}u_{n})=\partial_{x}(u_{n}v_{n}).
\label{DS*}
\end{equation}
(\ref{dds1}) and (\ref{DS*}) is nothing but (\ref{dd1}).
In the same way we can get (\ref{dd2})
using the other second flow. 
 This proves that the ALH  is the BT of the DS system with the constraints (\ref{cc}).
 
\setzero 
\section{Double KP System}

We first summarize  some background information on the Kadomtsev-Petviashvili
 (KP) hierarchy 
and ghost symmetries.
We use the Sato formalism of pseudo-differential operator 
calculus to describe  KP type integrable hierarchies 
of integrable nonlinear evolution equations about  
  the KP times $(t)=(t_{1}\equiv x,t_{2},\cdots)$:
\begin{equation}
{\cal L}=D +\sum_{i=1}^{\infty}p_{i}\partial ^{-i},
\;\;\;\;
\frac{\partial }{\partial t_{l}}{\cal L}=
[({\cal L})^{l}_{+}, {\cal L}],\;\;\;l=1,2,\cdots,
\label{kp1}
\end{equation}
where 
$D$ stands for the differential operator  $\partial/\partial x$,  and
 the subscripts  $+$ and $-$
of any pseudo-differential operator $A=\sum_{j}a_{j}D^{j}$
denote its purely differential part ($A_{+}=\sum_{j\geq 0}a_{j}D^{j}$)
and  its  
purely pseudo-differential part ($A_{-}=\sum_{j< 0}a_{j}D^{j}$)
respectively.

In the present approach 
eigenfunction (EF) $\Phi(t)$ and adjoint eigenfunction 
(adj EF) $\Psi(t)$ which satisfy the following equations play 
crucial role, 
\begin{equation}
\frac{\partial}{\partial t_{k}}\Phi=
({\cal L})^{k}_{+}(\Phi),
\;\;\;\;\;
\frac{\partial}{\partial t_{k}}\Psi=
({\cal L}^{*})^{k}_{+}(\Psi),
\label{EF}
\end{equation}
where ${\cal L}^{*}$ is the adjoint operator of ${\cal L}$.
Note that the above eigenfunctions $\Phi$ and $\Psi$
do not need to be the Baker-Akhiezer eigenfunctions.

A ghost symmetry is defined through an action of a vector field 
$\hat{\partial }_{\alpha}$ on the KP Lax operator
\begin{equation}
\hat{\partial }_{\alpha}{\cal L}=[{\cal M}_{\alpha}, {\cal L}]
,\;\;\;
{\cal M}_{\alpha}\equiv \sum_{a\in \{\alpha\}}
\Phi_{a}D^{-1}\Psi_{a},
\label{prop1}
\end{equation}
where $(\Phi_{a},\Psi_{a})_{a\in \{\alpha\}}$ are the 
some set of functions indexed by $\{\alpha\}$.
Commutativity of $\hat{\partial}_{\alpha}$ with
 $\partial_{l}$ implies that $(\Phi_{a},\Psi_{a})_{a\in \{\alpha\}}$
is a set of pairs of (adj-)EFs for  ${\cal L}$.

We now proceed to give an explicit construction of the 
ghost KP hierarchy.
Consider  an  infinite system of independent (adj-)EFs 
$(\Phi_{j},\Psi_{j})_{j=1}^{\infty}$
 of ${\cal L}$ and define the following  set of 
the ghost symmetry flows
\begin{equation}
\frac{\partial}{\partial \bar{t}_{s}}{\cal L}=
[{\cal M}_{s},{\cal L}],
\;\;\;\;
{\cal M}_{s}=\sum_{j=1}^{s}\Phi_{s-j+1}\partial^{-1}\Psi_{j},
\label{m}
\end{equation}
where $s=1,2,\cdots$.
Note that the ghost symmetry flows $\frac{\partial }{\partial \bar{t}_{s}}$ 
do commute.

To define the two Lax operators 
we introduce the Wronskian type determinant
\begin{equation}
W_{k}\equiv W_{k}[\Phi_{1},\cdots,\Phi_{k}]={\rm det}
||\partial ^{\alpha-1}\Phi_{\beta}||,\;\;\;
{\cal W}_{k}\equiv {\cal W}_{k}[\Psi_{1},\cdots,\Psi_{k}]={\rm \det}
||\partial ^{\alpha-1}\Psi_{\beta}||,
\end{equation}
where $\alpha,\beta=1,2,\cdots,k$.

We define the initial and the ghost Lax operators,
\begin{equation}
{\cal L}=\partial+\sum_{k=1}^{\infty}a_{k}
(D -\partial \ln\frac{{\cal W}_{k+1}}{{\cal W}_{k}})^
{-1}
\cdots
(D - \partial\ln\frac{{\cal W}_{2}}{{\cal W}_{1}})^
{-1},
\end{equation}
and 
\begin{equation}
\bar{{\cal L}}=\bar{\partial}+\sum_{k=1}^{\infty}b_{k}
(\bar{D} +\bar{\partial} \ln\frac{W_{k+1}}{W_{k}})^
{-1}
\cdots
(\bar{D} +\bar{\partial} \ln\frac{W_{2}}{W_{1}})^
{-1},
\end{equation}
where $\bar{\partial}=\partial _{\bar{t}_{1}}$ and 
hereafter we define $y=-\bar{t}_{1}$.

Let us introduce the non-standard orbit of successive Darboux-B\"{a}cklund 
(DB)  transformations
for the initial KP system
\begin{eqnarray}
& &{\cal L}(n+1)=T_{1}(n){\cal L}(n)T_{1}^{-1}(n),
\;\;\;\;
T_{1}(n)=\Phi_{1}D\Phi_{1}^{-1}\equiv
\Phi_{1}^{(n)}D(\Phi_{1}^{(n)})^{-1},
\nonumber \\
& &{\cal L}(n-1)=\hat{T}_{1}(n){\cal L}(n)\hat{T}_{1}^{-1}(n),
\;\;\;\;
\hat{T}_{1}(n)=\Psi_{1}D\Psi_{1}^{-1}\equiv
\Psi_{1}^{(n)}D(\Psi_{1}^{(n)})^{-1}.
\end{eqnarray}

The DB transformations for the ghost Lax operator is
\begin{equation}
\bar{{\cal L}}(n+1)=(\frac{1}{\Phi_{1}^{(n+1)}}\bar{D}^{-1}
\Phi_{1}^{(n+1)})
\bar{{\cal L}}(n)
(\frac{1}{\Phi_{1}^{(n+1)}}\bar{D}
\Phi_{1}^{(n+1)}).
\end{equation}

We can construct an infinite  set  of (adj-)EFs 
$(\bar{\Phi}_{j},\bar{\Psi}_{j})_{j=1}^{\infty}$
 for the ghost Lax operator $\bar{{\cal L}}$.
We can obtain the first relations 
\begin{equation}
\bar{\Phi}_{1}^{(n)}=\Psi_{1}^{(n+1)},
\;\;\;
\bar{\Psi}_{1}^{(n)}=\Psi_{1}^{(n+1)}.
\label{gef}
\end{equation}

Both Lax operators  ${\cal L}$ and $\bar{{\cal L}}$
define ``double KP system'',
\begin{eqnarray}
& &
\frac{\partial }{\partial t_{r}}{\cal L}
=[({\cal L}^{r})_{+},{\cal L}],\;\;\;\;
\frac{\partial }{\partial \bar{t}_{s}}{\cal L}
=[{\cal M}_{s},{\cal L}],
\nonumber \\
& &
\frac{\partial }{\partial \bar{t}_{s}}\bar{{\cal L}}
=[(\bar{{\cal L}}^{s})_{+},\bar{{\cal L}}],\;\;\;\;
\frac{\partial }{\partial \bar{t}_{r}}\bar{{\cal L}}
=[\bar{{\cal M}}_{s},\bar{{\cal L}}],
\label{prop3}
\end{eqnarray}
where 
$\bar{{\cal M}}_{r}$ is  defined in terms of the  ${\cal L}$  (adj)EFs
:$\bar{{\cal M}}_{r}=\sum_{i=1}^{r}\bar{\Phi}_{r-i+1}\bar{\partial}^{-1}
\bar{\Psi}_{i}$.

For $k=2$ (\ref{EF}) gives 

\begin{equation}
\frac{\partial \Phi_{1}}{\partial t_{2}}=
\frac{\partial ^{2}\Phi_{1}}{\partial x^{2}}+2p_{1}\Phi_{1},
\;\;\;
\frac{\partial \Psi_{1}}{\partial t_{2}}=
-\frac{\partial ^{2}\Psi_{1}}{\partial x^{2}}-2p_{1}\Phi_{1},
\label{k=2}
\end{equation}
while for $k=3$ we obtain

\begin{eqnarray}
& &\frac{\partial \Phi_{1}}{\partial t_{3}}=
\frac{\partial ^{3}\Phi_{1}}{\partial x^{3}}+3p_{1}
\frac{\partial \Phi_{1}}{\partial x}+
3p_{2}\Phi_{1}+
3\frac{\partial p_{1}}{\partial x}\Phi_{1},
\nonumber \\
& &\frac{\partial \Psi_{1}}{\partial t_{3}}=
\frac{\partial ^{3}\Psi_{1}}{\partial x^{3}}+3p_{1}
\frac{\partial \Psi_{1}}{\partial x}-
3p_{2}\Psi_{1}.
\label{k=3}
\end{eqnarray}

On the other hand (\ref{m}) gives
\begin{equation}
\frac{\partial p_{1}}{\partial y}=
\frac{\partial }{\partial x}(\Phi_{1}\Psi_{1}),
\label{m2}
\end{equation}
If we set $t_{2}=t$,
(\ref{k=2}) and (\ref{m2}) are nothing but the tally of the DS system  
(\ref{dd1}).

{From} (\ref{m}) we can obtain
\begin{equation}
\frac{\partial p_{2}}{\partial y}
=-\frac{\partial }{\partial x}
(\Phi_{1}\frac{\partial \Psi_{1}}{\partial x}).
\label{dd4}
\end{equation}
 
Using    (\ref{dd4}),
(\ref{k=3}) gives 
\begin{eqnarray}
& &\frac{\partial \Phi_{1}}{\partial t_{3}}=
\frac{\partial ^{3}\Phi_{1}}{\partial x^{3}}+
3(\frac{\partial }{\partial x}\int^{y}\Phi_{1}\Psi_{1}{\rm d}y)
\frac{\partial \Phi_{1}}{\partial x}
+
3[\int^{y}(\frac{\partial \Phi_{1}}{\partial x}
\frac{\partial \Psi_{1}}{\partial x}
+
\frac{\partial^{2} \Phi_{1}}{\partial x^{2}}\Psi_{1}){\rm d}y]
\Phi_{1},
\nonumber \\
& &\frac{\partial \Psi_{1}}{\partial t_{3}}=
\frac{\partial ^{3}\Psi_{1}}{\partial x^{3}}+
3(\frac{\partial }{\partial x}\int^{y}\Phi_{1}\Psi_{1}{\rm d}y)
\frac{\partial \Psi_{1}}{\partial x}
+
3[\int^{y}(\frac{\partial \Phi_{1}}{\partial x}
\frac{\partial \Psi_{1}}{\partial x}
+
\Phi_{1}\frac{\partial^{2} \Psi_{1}}{\partial x^{2}}){\rm d}y]
\Psi_{1},
\end{eqnarray}
This is the tally of the higher DS system.
If we set $x=y$
we can obtain 
the complex modified  Korteweg-de Vries (cmKdV) equation
which is the first higher order equations in the   
nonlinear Schr\"{o}dinger  hierarchy.

According to (\ref{prop3}) and (\ref{prop1}) there exists
a duality mapping between the double KP systems  of 
(\ref{prop3}) defined by ${\cal L}$ and $\bar{\cal L}$, respectively 
under the change  $(t)\leftrightarrow (\bar{t})$, 
$\Phi\leftrightarrow\bar{\Phi}$ and $\Psi\leftrightarrow\bar{\Psi}$.
Then for example we can get the equations by the duality mapping of 
(\ref{k=2}) and (\ref{m2})
\begin{equation}
\frac{\partial \bar{\Phi}_{1}}{\partial \bar{t}_{2}}=
\frac{\partial ^{2}\bar{\Phi}_{1}}{\partial y^{2}}+2q_{1}\bar{\Phi}_{1},
\;\;\;
\frac{\partial \bar{\Psi}_{1}}{\partial \bar{t}_{2}}=
-\frac{\partial ^{2}\bar{\Psi}_{1}}{\partial y^{2}}-2q_{1}\bar{\Psi}_{1},
\label{k=2*}
\end{equation}
and 
\begin{equation}
\frac{\partial q_{1}}{\partial x}=
\frac{\partial }{\partial y}(\bar{\Phi}_{1}\bar{\Psi}_{1}).
\label{m2*}
\end{equation}
 
Using (\ref{gef}) for any site $n$ on the DB-orbit 
we can obtain 
\begin{equation}
\frac{\partial \Phi_{1}}{\partial \bar{t}_{2}}=
-\frac{\partial ^{2}\Phi_{1}}{\partial y^{2}}-2q_{1}\Phi_{1},
\;\;\;
\frac{\partial\Psi_{1}}{\partial \bar{t}_{2}}=
\frac{\partial ^{2}\Psi_{1}}{\partial y^{2}}+2q_{1}\Psi_{1},
\label{k=2**}
\end{equation}
and 
\begin{equation}
\frac{\partial q_{1}}{\partial x}=
\frac{\partial }{\partial y}(\Phi_{1}\Psi_{1}),
\label{m2**}
\end{equation}
where we rewrite $\bar{p}_{1}=q_{1}$.
If we set $\tau=-\bar{t}_{2}$, these equations are
nothing but the tally of the DS system 
(\ref{dd2}).
In the same way we can obtain the tally of the higher 
DS system.
If we set $y=\bar{t}_{1}$, we can obtain the other type 
 (the sign is ``$-$'' in (\ref{dd1}) and (\ref{dd2}))  system.
Note that these equations have  dark type 
  solutions. 

{From} these results 
we can obtain  one  hierarchy 
from the flows with respect to  $(t)$  and $y$ of double KP system.
On  the other hand we can get  the other hierarchy
from the flows  with respect to   $(-\bar{t})$ and $x$.
These two hierarchies are compatible and 
any linear combination of them is integrable.
The couple of these hierarchies are 
the DS system.
Note that  dependent variables of DS system 
is (adj-)EF  of duble KP system.

\section{Coupled Derivative Nonlinear Schr\"{o}dinger equations}
If we set $x=y$
in  (\ref{ce1}) and (\ref{ce2}),  
we can obtain 
\begin{eqnarray}
-\partial_{x}u_{n}^{(i)}=u_{n+1}^{(i)}+
(u_{n}^{(1)}v_{n+1}^{(1)}+u_{n}^{(2)}v_{n+1}^{(2)})u_{n}^{(i)},
\nonumber \\
\partial_{x}v_{n}^{(i)}=v_{n-1}^{(i)}+
(u_{n-1}^{(1)}v_{n}^{(1)}+u_{n-1}^{(2)}v_{n}^{(2)})v_{n}^{(i)},
\label{ce}
\end{eqnarray}
\nonumber \\
where $i=1,2$.
(\ref{ce}) are  the BT of the coupled NLS equations (\ref{cnls}).

Let us consider the following transformation,
\begin{equation}
U_{n}^{(i)}=u_{n}^{(i)},\;\;\;\;
V_{n}^{(i)}=v_{n+1}^{(i)},\;\;\; i=1,2.
\label{VT}
\end{equation}
This transformation is obviously invertible.
In the new variables chain equations (\ref{ce})
 take the form 
\begin{eqnarray}
-\partial_{x}U_{n}^{(i)}=U_{n+1}^{(i)}+(U_{n}^{(1)}V_{n}^{(1)}
+U_{n}^{(2)}V_{n}^{(2)})U_{n}^{(i)},
\nonumber \\
\partial_{x}V_{n}^{(i)}=V_{n-1}^{(i)}+(U_{n}^{(1)}V_{n}^{(1)}
+U_{n}^{(2)}V_{n}^{(2)})V_{n}^{(i)},
\label{14}
\end{eqnarray}
where 
 $i=1,2$.

It follows from the above system of equations  that variables 
$u_{n}^{(i)}, v_{n}^{(i)},$ can be expressed  in terms of 
$U_{n}^{(i)}, V_{n}^{(i)},$ 
\begin{equation}
u_{n}^{(i)}=U_{n}^{(i)},\;\;\;\;
v_{n}^{(i)}=V_{n-1}^{(i)}=\partial_{x}V_{n}^{(i)}-
(U_{n}^{(1)}V_{n}^{(1)}+U_{n}^{(2)}V_{n}^{(2)})V_{n}^{(i)},
\;\;\;i=1,2.
\label{15}
\end{equation}
This defines a Miura transformation (MT)
\begin{equation}
u^{(i)}=U^{(i)},\;\;\;v^{(i)}=\partial_{x}V^{(i)}-
(U^{(1)}V^{(1)}+U^{(2)}V^{(2)})V^{(i)},\;\;\;i=1,2.
\label{MT}
\end{equation}
To find the transformed equations let us rewrite 
the system of (\ref{cnls}) in the new variables 
$U^{(i)},V^{(i)}$.
Hereafter  we only consider the bright case, that is,  the sign is ``$+$''
 in (\ref{cnls}). 
It follows from (\ref{VT}) that
\begin{eqnarray}
\partial_{t}U_{n}^{(i)}&=&\partial^{2}_{x}U_{n}^{(i)}
+2(U_{n}^{(1)}V_{n-1}^{(1)}+U_{n}^{(2)}V_{n-1}^{(2)})U_{n}^{(i)},
\nonumber \\
-\partial_{t}V_{n}^{(i)}&=&\partial^{2}_{x}V_{n}^{(i)}
+2(U_{n+1}^{(1)}V_{n}^{(1)}+U_{n+1}^{(2)}V_{n}^{(2)})V_{n}^{(i)},
\;\;\;i=1,2.
\end{eqnarray}
The variables $V_{n}^{(i)}$ is already expressed in terms of 
$U_{n}^{(i)}, V_{n}^{(i)}$ (\ref{15}).
It follows from (\ref{14}) that
\begin{equation}
U_{n+1}^{(i)}=-\partial_{x}U_{n}^{(i)}-
(U_{n}^{(1)}V_{n}^{(1)}+U^{(2)}_{n}V_{n}
^{(2)})U_{n}^{(i)}.
\label{U}
\end{equation}
Thus the transformed equations by the MT
(\ref{MT}) of the system (\ref{cnls}) is of the form 
\begin{eqnarray}
\partial_{t} U^{(i)}
&=&
\partial^{2}_{x}U^{(i)}+2(U^{(1)}\partial_{x}V^{(1)}+U^{(2)}
\partial_{x}V^{(2)})U^{(i)}
-2(U^{(1)}V^{(1)}+U^{(1)}V^{(1)})^{2}U^{(i)},
\nonumber \\
-
\partial_{t} V^{(i)}
&=&
\partial^{2}_{x}V^{(i)}-2(V^{(1)}\partial_{x}U^{(1)}+V^{(2)}
\partial_{x}U^{(2)})U^{(i)}
-2(U^{(1)}V^{(1)}+U^{(1)}V^{(1)})^{2}V^{(i)},
\label{cdnls}
\end{eqnarray}
where $i=1,2$. 
(\ref{cdnls}) are coupled derivative nonlinear Schr\"{o}dinger 
(DNLS) equations. 
There are two MT between (\ref{cnls}) and (\ref{cdnls}).
To construct the other MT 
 we expressed  variables 
$u^{(i)}=u_{n+1}^{(i)}, v^{(i)}=v_{n+1}^{(i)}$
in terms of $U^{(i)}=U_{n}^{(i)}, V^{(i)}=V_{n}^{(i)}$.
It follows from (\ref{VT}) $u_{n+1}^{(i)}=U_{n+1}^{(i)}$,
$v_{n+1}^{(i)}=V_{n}^{(i)}$ and from (\ref{U})
that
\begin{equation}
u^{(i)}=-U_{x}^{(i)}-(U^{(1)}V^{(1)}+U^{(1)}V^{(1)})U^{(i)},
\;\;\;
v^{(i)}=V^{(i)}.
\label{MT1}
\end{equation}
This is the second MT
 linking 
(\ref{cnls}) and (\ref{cdnls}).

{From} (\ref{VT}),  $u_{n+1}^{(i)}=U_{n+1}^{(i)}$ and 
$v_{n+1}^{(i)}=V_{n}^{(i)}$  
we can conclude that 
the difference between the 
 coupled NLS equation  and the coupled DNLS equations 
are  the complex conjugation between $\{u_{n}^{(i)}\}$ and 
$\{v_{n}^{(i)}\}$.
For  the coupled NLS equations  
$u_{n}^{(i)}$ and $v_{n}^{(i)}$ are  complex conjugate.
On the other hand 
for  the coupled DNLS equations 
$u_{n}^{(i)}$ and $v_{n+1}^{(i)}$ are  complex conjugate.

\newcommand{\eqref}[1]{eq.(\ref{#1})}
\newcommand{\secref}[1]{\ref{#1}}
\newcommand{\beq}{\begin{equation}}
\newcommand{\eeq}{\end{equation}}
\newcommand{\bl}[1]{\makebox[#1em]{}}
\newcommand{\pa}{\partial}
\newcommand{\ovl}[1]{\overline{#1}}
\newcommand{\ul}[1]{\underline{#1}}

\newcommand{\qtil}{\hat{q}}
\newcommand{\Qtil}{\hat{Q}}
\newcommand{\ftil}{\tilde{f}}
\newcommand{\gtil}{\tilde{g}}
\newcommand{\htil}{\tilde{h}}
\newcommand{\uhat}{\hat{u}}
\newcommand{\vhat}{\hat{v}}

\newcommand{\ee}{\mbox{e}}

Here we consider the coupled version of the Chen-Lee-Liu (CLL) type equations
\cite{DNLS2},
\cite{Ka}
\beq
\ii \Qtil_T^{(i)} = \Qtil_{XX}^{(i)} 
 + \ii\alpha  (\sum_{k}^{N}|\Qtil^{(k)}|^{2})\Qtil^{(i)}_X ,
\label{hc}
\eeq
If we set
\beq
Q^{(i)}=\Qtil ^{(i)}\exp\left( -2\ii\delta\sum_{k}^{N}
\int^{X}|\Qtil^{(k)}|^2 \mbox{d}X \right),
\label{eqn:GT}
\eeq
then (\ref{hc}) is gauge-equivalent \cite{WS}\cite{h3} to
\begin{equation}
\ii Q _T ^{(i)}=Q _{XX}^{(i)} 
-2\ii\delta AQ^{(i)}+2\ii\delta BQ^{(i)}+\ii(4\delta+\alpha)
\rho^{2}_{Q}Q^{(i)}_{X}
 + \delta(4\delta-\alpha)\rho^{4}_{Q}  Q^{(i)} ,
\label{eqn:GDNLS0}
\end{equation}
where
\begin{eqnarray}
A&=&\sum_{k}^{N}(Q_{X}^{(k)}Q^{(k)*}-Q^{(k)*}_{X}Q^{(k)}),\;\;\;
B=\sum_{k}^{N}(Q_{X}^{(k)}Q^{(k)*}+Q^{(k)*}_{X}Q^{(k)}),
\nonumber \\
\rho^{2}_{Q}&=&\sum_{(k)}^{N}|Q^{(k)}|^2.
\end{eqnarray}
(\ref{eqn:GDNLS0}) are  the new coupled version of the generalized 
coupled DNLS  equations.
(\ref{eqn:GDNLS0}) is different from the the generalized 
coupled DNLS  equations  which we obtain in the previous 
works.\cite{h1},\cite{h2}
These equations are the coupled version 
of the Kaup-Newell equation,\cite{DNLS1}
\beq
\ii \Qtil_T^{(i)} = \Qtil_{XX}^{(i)} 
 + \ii\alpha  [(\sum_{k}^{N}|\Qtil^{(k)}|^{2})\Qtil^{(i)}]_X ,
\label{hc1}
\eeq
They   do  not have  MT to CNLS equations.

If we set in (\ref{cdnls})
\begin{eqnarray}
& &1)\;\;x=\ii X,\;\;t=-\ii T,\;\;Q^{(i)}=U^{(i)},\;\;Q^{(i)*}=V^{(i)},\;\;
\alpha=-4\delta=2,
\nonumber \\
& &2)\;\;x=\ii X,\;\;t=\ii T,\;\;Q^i=V^{(i)},\;\;Q^{(i)*}=U^{(i)},\;\;
-\alpha=4\delta=2,
\end{eqnarray}
then (\ref{cdnls}) is  gauge equivalent   to (\ref{eqn:GDNLS0}).

\setzero
\section{Concluding Remarks}

We  have considered  the (coupled) Davey-Stewartson (DS) system and 
its B\"{a}cklund transformations (BT).
Relations  among   the DS   system, the double KP system  and 
 the Ablowitz-Ladik hierarchy (ALH) have been  established.

The double KP system has two sets of times, 
the original $(t)\equiv(t_{1}\equiv x,t_{2},\cdots)$ 
and the ghost ones ($\bar{t})\equiv(\bar{t}_{1}\equiv -y,\bar{t}_{2},\cdots)$.
Then we can obtain one  hierarchy 
about the flows to $y$ and ($t)$.
We can get the other one from   
the flows to $x$ and $(-\bar{t})$.
The first equations 
in each hierarchy are DS equations (\ref{dd1}) and (\ref{dd2}).
 These two hierarchies are compatible and 
any linear combination of them is   integrable.
The couple of these hierarchies are 
the DS system.
Note that  dependent variables of DS system 
is (adj-)EF  of double KP system.

The ALH has been found to be  the BT of  the DS  system 
in the case $t_{k}=\bar{t}_{k}^{*}$ and $u_{n}=v_{n}^{*}$. 
$u$ and $v$ are dependent variables 
of DS system.
The asterisk is  conjugation.
The flows  of ALH are also those 
of double KP and DS systems.
For this constraints 
the first equations which contain no discrete variables $n$
are  the complex sine-Goldon  equations  instead of the DS equations.\cite{V3}

 {From} the BT of the coupled DS equations we  can obtain the new coupled 
derivative nonlinear  Schr\"{o}dinger (DLNS) equations. 
We can obtain the Miura transformations (MT)
between   the coupled NLS and the coupled DNLS equations.
It is the new type  of the coupled DNLS equations.
The other coupled DNLS equations  do  not have  the MT to the  CNLS equations.
We will report on 
the relations between these two coupled DNLS equations and 
 investigations  about these equations  in next paper.

\end{document}